\documentclass[
reprint, amsmath,amssymb,aps,prb,twocolumn]{revtex4-2}
\usepackage[force]{feynmp-auto}
\usepackage{subfigure}
\usepackage{graphicx}
\usepackage{float}
\usepackage{dcolumn}
\usepackage{bm}
\usepackage{physics}
\usepackage[backref]{hyperref}
\usepackage{mathrsfs}

\begin{document}
	
	\preprint{APS/123-QED}
\title{Quantum phase transition in a 2D atomic Bose gas with a \textit{g}-wave Feshbach resonance}
\author{Fan Zhang}
\email{rapanuifz@gmail.com}
\affiliation{Hefei National Laboratory, Hefei 230088, China}
\author{Lan Yin}
\affiliation{School of Physics, Peking University, Beijing 100871, China}
\date{\today}
	\begin{abstract}
		Recent repoet on the formation of two-dimensional Bose-Einstein condensates (BECs) of spinning \textit{g}-wave molecules is surprise. Here we study quantum phase transition in the quasi-2D atomic Bose gas with a \textit{g}-wave Feshbach resonance, and show that there are two phase transitions in this system: from a phase with only a atomic Bose-Einstein condensate to a phase with coexistence of atomic and molecular condensation and from the coexistence phase to a phase with only a molecular Bose-Einstein condensate. We show that the \textit{g}-wave resonance Feshbach has two features in the phase transitions, a phase with only a atomic Bose-Einstein condensate is no longer forbidden and resonance effects appear only at levels beyond the mean-field. We determine the $T=0$ beyond mean-field phase diagram of the gas as a function of magnetic field and molecular condensation density. We also detemine the ratio of atomic to molecular condensation in the coexistence phase, which can be tested in experiments.
	\end{abstract}
	\maketitle
	Recently, Cheng Chin et al. reported the formation of two-dimensional Bose-Einstein condensates (BECs) of spinning \textit{g}-wave molecules\cite{zhang2020atomic}. These molecules are created through the \textit{g}-wave Feshbach resonance. Here we study theoretically the quantum phase transition in the 2D atomic Bose gas with a \textit{g}-wave Feshbach resonance.
	
	We begin with a quasi-2D grand-canonical Hamiltonian of spinless bosonic atoms, resonantly interacting through a diatomic molecular state with a large orbital angular momentum $l = 4$ and projection along the molecular axis $m_l = 2$:
\begin{eqnarray}
&\hat{H}_{\mu}=\sum\limits_{\vb p\sigma}(\varepsilon_{\vb p\sigma}-\mu_{\sigma})a^{\dagger}_{\vb p\sigma}a_{\vb p\sigma} \label{ham}\\ \notag
&+\frac{1}{2V}\sum\limits_{\vb p, \vb p', \vb q, \sigma}{g_{\sigma}}a^{\dagger}_{\vb{(p+q)}\sigma}a^{\dagger}_{\vb{(p'-q)}\sigma}a_{\vb{p}\sigma}a_{\vb{p'}\sigma}\\ \notag
&+\frac{g_{12}}{V}\sum\limits_{\vb p, \vb p', \vb q}a^{\dagger}_{\vb{(p+q)}1}a^{\dagger}_{\vb{(p'-q)}2}a_{\vb{p}2}a_{\vb{p'}1}\\ \notag
&+\frac{g}{\sqrt{V}}\sum\limits_{\vb{q,p}}[j_4(\abs{\vb {\tilde r_0}}\abs{\vb p})Y^*_{42}(\hat{\vb p})a_{\vb q2}a^{\dag}_{(\vb{p+\frac{q}{2}})1}a^{\dag}_{(\vb{-p+\frac{q}{2}})1}+h.c.]. 
\end{eqnarray}
    Here $a^{\dagger}_{\vb p\sigma}, a_{\vb p\sigma}$ are bosonic field operators for atoms ($\sigma=1$) with kinetic energy $\varepsilon_{\vb p1}=\frac{\vb{p}^2}{2m}$ and molecules ($\sigma=2$) with kinetic energy $\varepsilon_{\vb p2}=\frac{\vb{p}^2}{4m}$. The range of a one-dimensional box potential is $z\in[-L, L]$ with periodic boundary conditions, and the momentum at $z$ is $p_z = n\pi/L (n = 0, \pm1, ...)$. $\mu_1=\mu$ and $\mu_2=2\mu-\nu$ are effective chemical potentials. The detuning $\nu$ is the energy of a bare molecule at rest, that can be experimentally controlled with an external magnetic field. $g_1,\, g_2$ and $g_{12}$ describe the interatomic, intermolecular and atomic-molecular background scattering intensities respectively. In the Feshbash resonance term $g$ is the coupling constant, $Y_{42}$ is the spherical harmonic, which is anisotropic and $j_4$ is the spherical Bessel function. The order of magnitude of $\abs{\vb {\tilde r_0}}$ is $\mathrm{nm}$, which is the size of molecules. The $j_4$ function is approximately $p^4$ in small $p$ ($p<\frac{1}{\tilde r_0}$) and converges as $p$ becomes larger. This Hamiltonian is a generalization of the \textit{s}-wave two-channel model, in which the Feshbash resonance term ($a_{\vb q2}a^{\dag}_{(\vb{p+\frac{q}{2}})1}a^{\dag}_{(\vb{-p+\frac{q}{2}})1}$) is proportional to a constant $g$\cite{drummond1998coherent,timmermans1999feshbach,kokkelmans2004degenerate}. We will see how this distinction makes the phase transition with Feshbash resonances with non-zero momentum distinct from the \textit{s}-wave case\cite{radzihovsky2004superfluid,radzihovsky2008superfluidity}. We show the derivation of (\ref{ham}) in the Supplementary Materials I.
    
    The order parameters in our problem are the atomic condensate wave function $\Psi_{10}\equiv \langle a_{\vb{p}=0,1} \rangle/\sqrt{V}$ and the molecular condensate wave function $\Psi_{20}\equiv \langle a_{\vb{p}=0,2} \rangle/\sqrt{V}$. There are four possible phases: (i) “Normal” (N): $\Psi_{10}=0, \Psi_{20}=0$, (ii) “atomic superfluid” (ASF): $\Psi_{10}\neq0, \Psi_{20}=0$, (iii) “molecular superfluid” (MSF): $\Psi_{10}=0, \Psi_{20}\neq0$ and (iv) “atomic-molecular superfluid” (AMSF): $\Psi_{10}\neq0, \Psi_{20}\neq0$. For a fixed chemical potentials $\mu$ and detuning $\nu$, the order parameters $\Psi_{\sigma 0}$ should minimize energy functional $H_{\mu}[\Psi_{\sigma 0}]$.
    
\begin{figure}[htbp]
    	\centering  
    	\subfigure[]{
    		\label{Fig.sub.1}
    		\includegraphics[width=0.23\textwidth]{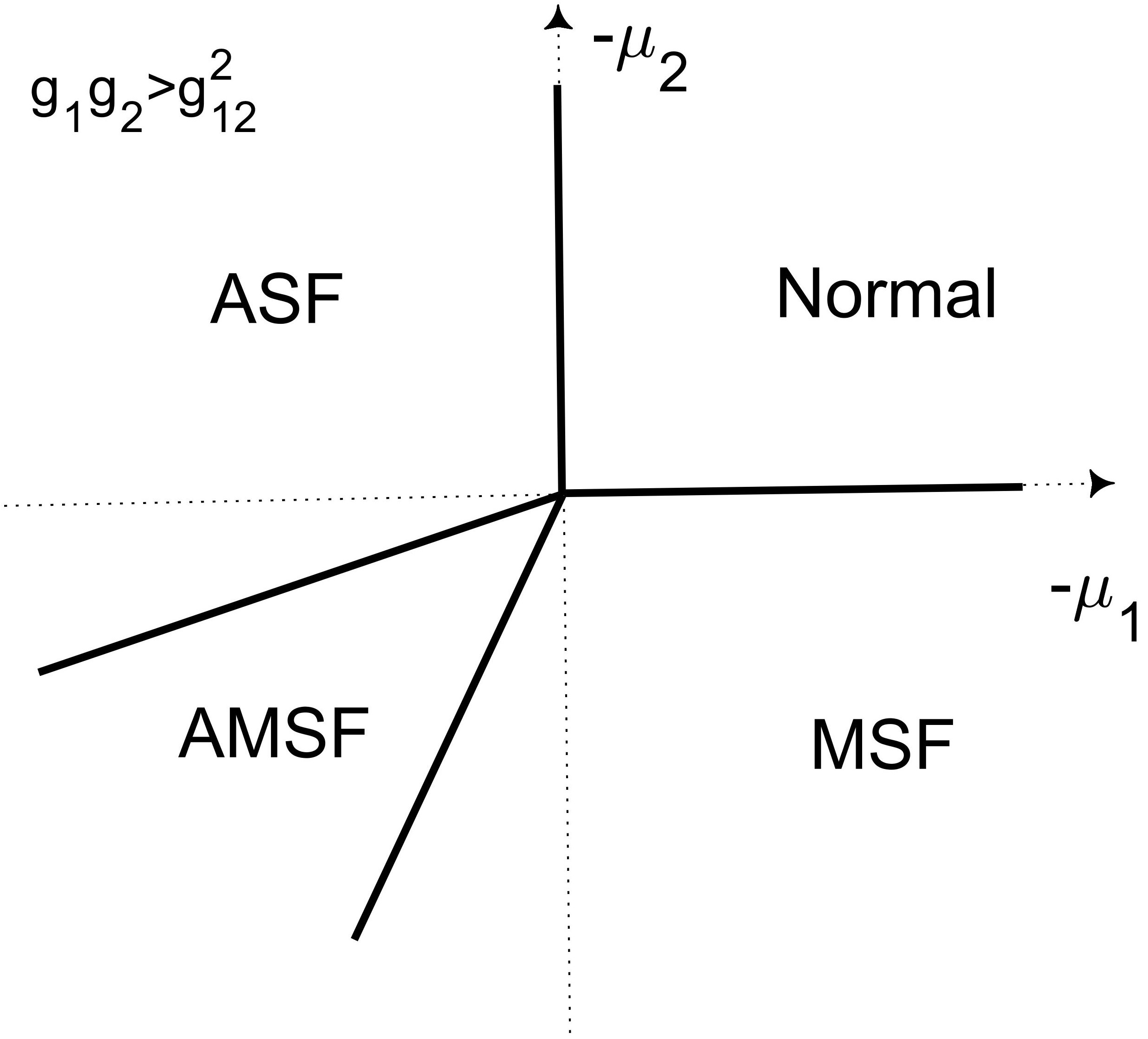}}
    	\subfigure[]{
    		\label{Fig.sub.2}
    		\includegraphics[width=0.23\textwidth]{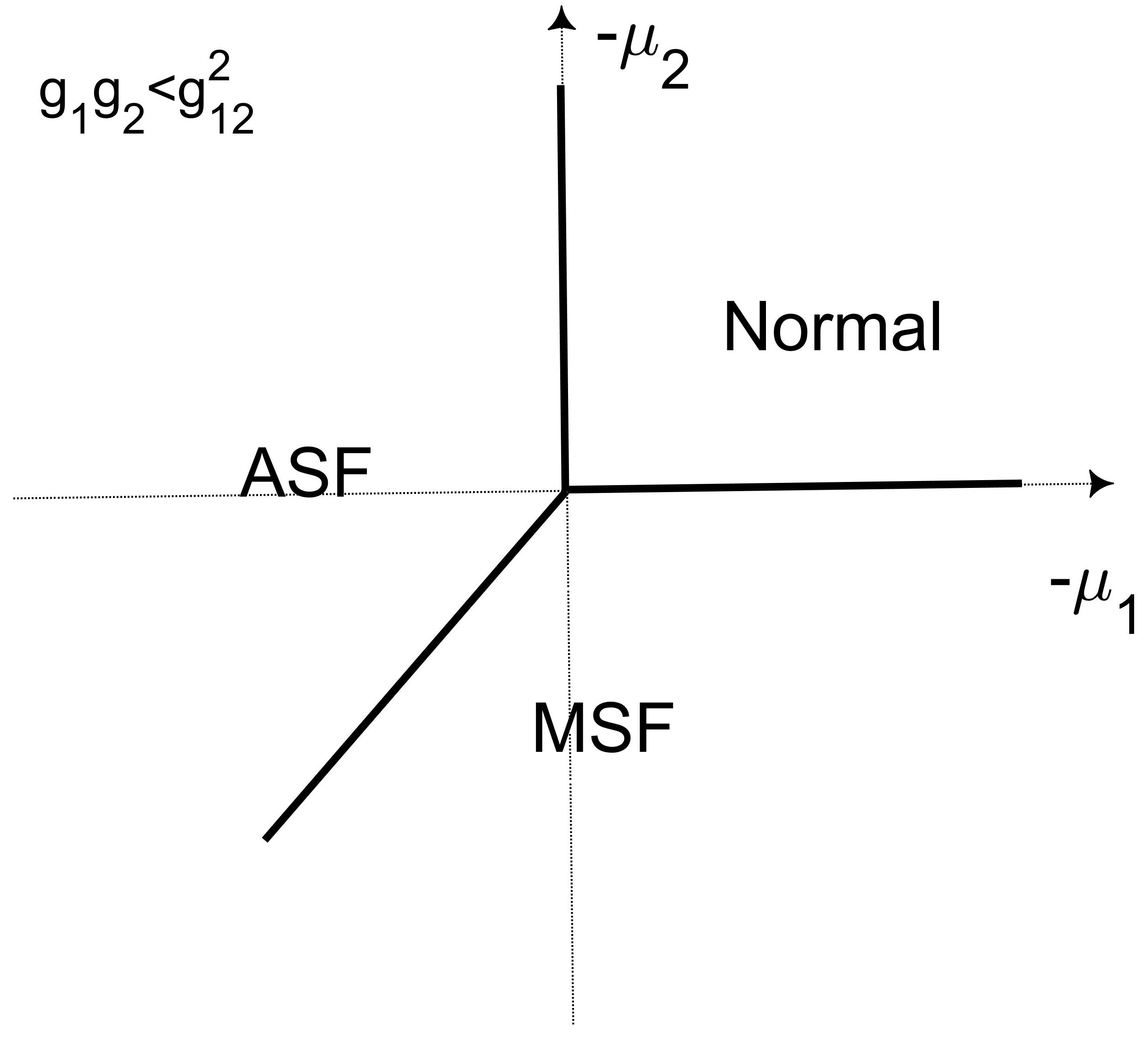}}
    
    	\caption{Mean-field phase diagram}
    	\label{mf}
    \end{figure}

    At the mean-field level, all fluctuations are ignored and the energy functional $H_{\mu}[\Psi_{\sigma 0}]\approx H_{mf}[\Psi_{\sigma 0}]$, in which one substitutes the order parameters for the field operators in (\ref{ham}). The mean-field energy density is
    \begin{eqnarray}
    	&e_{mf}= H_{mf}[\Psi_{\sigma 0}]/V \label{hmf} \\
    	&=\sum\limits_{\sigma}(-\mu_{\sigma}\abs{\Psi_{\sigma 0}}^2+\frac{g_{\sigma}}{2}\abs{\Psi_{\sigma 0}}^4)+g_{12}\abs{\Psi_{1 0}}^2\abs{\Psi_{2 0}}^2. \notag
    \end{eqnarray}
    One of the characteristics of high angular momentum Feshbach resonance is that the ASF phase is no longer forbidden. In the \textit{s}-wave case the resonance term in the mean-field energy functional hybridizes states of a pair of atoms and a molecule, and as a result a finite molecular condensate is always induced in a state where atoms are condensed, i.e. an equilibrium phase in which atoms are condensed, but molecules are not is forbidden\cite{romans2004quantum,radzihovsky2004superfluid,radzihovsky2008superfluidity}. Compared to the \textit{s}-wave case in the \textit{g}-wave case the resonance term vanishes in the mean-field energy functional due to $\vb p \to 0, j_4(\abs{\vb {\tilde r_0}}\abs{\vb p})\to 0$ causing the ASF phase to be allowed, and similar result is found for other high angular momentum Feshbach resonances. The mean-field phase diagram (Fig. \ref{mf}) as a function of $\mu_{1,2}$ can be worked out by minimizing (\ref{hmf}).
    
    When condensation occurs, the phase of the condensate wave function is fixed, i.e. global $U(1)$ symmetry is broken. According to the Goldstone's theorem\cite{liang1998j} this leads to a collective $gapless$ excitation mode, which corresponds to atom-like excitation in the ASF phase and molecular-like excitation in the MSF phase. To study the excitations of atomic and molecular superfluids, we make the Bogoliubov approximation\cite{pethick2008bose} to (\ref{ham}). We expand the field operators about the mean-field solution: $\hat{\Psi}_{\sigma}(\vb{r})=\Psi_{\sigma 0}+\hat{\phi}_{\sigma}(\vb{r})$, and keep terms in the Hamiltonian only to quadratic order in the small fluctuations $\hat{\phi}_{\sigma}(\vb{r})=\sum\limits_{\vb{p}}a_{\vb p\sigma}e^{i\vb{p}\cdot\vb{r}}/\sqrt{V}$ ($p_z = n\pi/L, n = 0, \pm1, ...$). The Bogoliubov Hamiltonian is $\hat{H}_{Bog}=H_{mf}[\Psi_{\sigma 0}]+\hat{H}_2$. One obtains 
    \begin{eqnarray}
    	&\hat{H}_2=\sum\limits_{\vb p\neq0, \sigma}[\tilde{\varepsilon}_{\vb p, \sigma}a^{\dagger}_{\vb p, \sigma}a_{\vb p, \sigma}+\frac{1}{2}(\lambda_{\sigma}a^{\dagger}_{\vb p, \sigma}a^{\dagger}_{-\vb p, \sigma}+h.c.)]\\
    	&+\sum\limits_{\vb p\neq0}(t_1a^{\dagger}_{\vb p1}a_{\vb p2}+t_2a^{\dagger}_{\vb p1}a^{\dagger}_{-\vb p2}+h.c.), \notag
    \end{eqnarray}
    where $\vb p=(n\pi/L)\cdot \hat{\vb p}_z+\vb{p_{\parallel}}, (n=0,\pm1,...)$ and the coefficients are given by
    \begin{eqnarray}
    	&\tilde{\varepsilon}_{\vb p1}=\varepsilon_{\vb p1}-\mu_1+2{g}_1\abs{\Psi_{10}}^2\\ \notag
    	&\tilde{\varepsilon}_{\vb p2}=\varepsilon_{\vb p2}-\mu_2+2{g}_2\abs{\Psi_{20}}^2\\ \notag
    	&\lambda_1={g}_1\Psi_{10}^2+2{g}j_4(\abs{\vb {\tilde r_0}}\abs{\vb p})Y_{42}^*(\hat{\vb p})\Psi_{20}, \lambda_2={g}_2\Psi_{20}^2\\ \notag
    	&t_1=2{g}j_4(\abs{\vb {\tilde r_0}}\abs{\frac{\vb p}{2}})Y_{42}^*(\frac{\hat {\vb p}}{2})\Psi_{10}^*, t_2=0.\\ \notag
    	&\varepsilon_{\vb p,\sigma}=[(n\pi/L)^2+(\vb{p_{\parallel}})^2]/(2\sigma m), (n=0,\pm1,...)\\ \notag
    \end{eqnarray}
    Here we negelect the background scattering $g_{12}$. The Bogoliubov transformation ${\gamma}_{\vb{p}\sigma}=u_{\vb{p}\sigma}a_{\vb{p}\sigma}+v_{\vb{p}\sigma}a^{\dagger}_{-\vb{p}\sigma}$ leads to:
    \begin{eqnarray}
    	&\hat{H}_2=\sum\limits_{\vb p\neq 0\sigma}E_{\vb{p}\sigma}\gamma^{\dagger}_{\vb{p}\sigma}\gamma_{\vb{p}\sigma}+\frac{1}{2}(E_{\vb{p}\sigma}-\tilde{\varepsilon}_{\vb p \sigma}) \label{h2}\\ 
	    &E_{\vb{p}\sigma}^2=e_{\vb{p}}\pm\sqrt{d_{\vb{p}}^2+c_{\vb{p}}^2} \label{ex}\\ \notag
	    &e_{\vb{p}}=\frac{1}{2}(\tilde{\varepsilon}_{\vb p1}^2-\abs{\lambda_1}^2+\tilde{\varepsilon}_{\vb p2}^2-\abs{\lambda_2}^2)+\abs{t_1}^2-\abs{t_2}^2 \\ \notag
	    &d_{\vb{p}}=\frac{1}{2}(\tilde{\varepsilon}_{\vb p1}^2-\abs{\lambda_1}^2-\tilde{\varepsilon}_{\vb p2}^2+\abs{\lambda_2}^2)\\ \notag
	    &c_{\vb{p}}^2=2\Re{\lambda_1^* \lambda_2 t_1^2}+2\tilde{\varepsilon}_{\vb p1} \tilde{\varepsilon}_{\vb p2} \abs{t_1}^2\\ \notag
	    &+ \tilde{\varepsilon}_{\vb p2}^2 \abs{t_1}^2 -\abs{t_1}^2 \abs{\lambda_2}^2 +\abs{t_1}^2 \tilde{\varepsilon}_{\vb p1}^2-\abs{t_1}^2 \abs{\lambda_1}^2.
\end{eqnarray}
        At $\vb{p}\to\vb{0}$ it is easy to verify that $t_1=0, c_{\vb{0}}=0, \lambda_1={g}_1\Psi_{10}^2\equiv n_{10}g_1, \lambda_2={g}_2\Psi_{20}^2\equiv n_{20}g_2$, in the ASF phase $\mu_1=n_{10}g_1$ the atom-like excitation $E^a_{\vb p}\to\sqrt{\tilde{\varepsilon}_{\vb p1}^2-\abs{\lambda_1}^2}=\sqrt{\varepsilon^2_{\vb p1}+2n_{10}g_1\varepsilon_{\vb p1}}$ is gapless, and in the MSF phase $\mu_2=n_{20}g_2$ the molecular-like excitation $E^m_{\vb p}\to\sqrt{\tilde{\varepsilon}_{\vb p2}^2-\abs{\lambda_2}^2}=\sqrt{\varepsilon^2_{\vb p2}+2n_{20}g_2\varepsilon_{\vb p2}}$ is gapless.
        For high angular momentum Feshbach resonances including the \textit{g}-wave case, the resonance term vanishes in the mean-field energy functional (\ref{hmf}) and appears in the Bogoliubov Hamiltonian (\ref{h2}), leading to anisotropic excitations (\ref{ex}) and contributing to the LHY energy\cite{huang1957quantum} (summation of the zero-point energy of the Bogoliubov excitation modes, which is the last summation term of (\ref{h2})), which will correct the energy functional and the phase diagram.
        
\begin{figure}[H] 
        	\centering 
        	\includegraphics[width=0.45\textwidth]{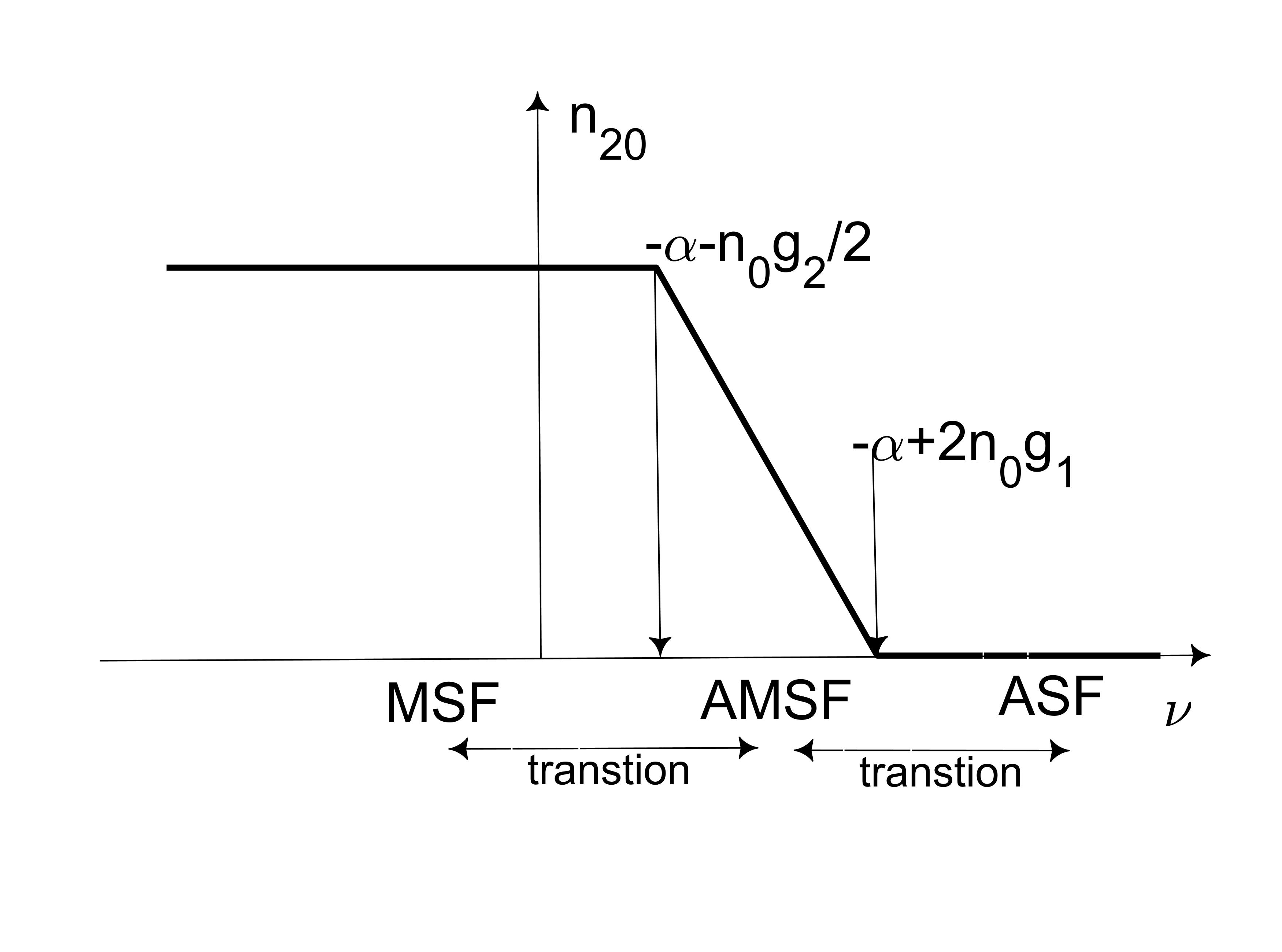} 
        	\caption{Schematic diagram of the ASF-AMSF-MSF phase transitions} 
        	\label{tran} 
        \end{figure}

        When higher order fluctuations are considered, the energy functional $H_{\mu}[\Psi_{\sigma 0}]$ is approximated as $H_{bmf}[\Psi_{\sigma 0}]$ and the beyond mean-field energy density is $e_{bmf}=H_{bmf}[\Psi_{\sigma 0}]/V=e_{mf}+e_{LHY}$. Taking into account the renormalization effects, the LHY energy density is
        \begin{eqnarray}
            e_{LHY}=\frac{1}{2V}\sum\limits_{\vb p\sigma}(E_{\vb{p}\sigma}-\tilde{\varepsilon}_{\vb p \sigma}+\frac{(n_{\sigma 0}g_{\sigma})^2}{2{\varepsilon}_{\vb p \sigma}}) 
        \end{eqnarray}
        The detailed calculation of the LHY energy of quasi-2D Feshbash resonance system is in the Supplementary Materials II. According to the mean-field phase diagram, molecular condensate is always 0 when $\mu_2<0$, but we find that finite molecular condensate always reduces the LHY energy. To obtain a more accurate phase diagram, we minimize the beyond mean-field energy density fuctional $e_{bmf}[\Psi_{\sigma 0}]$:
        \begin{eqnarray}
        	&0=\frac{\partial e_{bmf}}{\partial n_{10}}=-\mu_1+n_{10}g_1+\frac{\partial e_{LHY}}{\partial n_{10}} \\
        	&0=\frac{\partial e_{bmf}}{\partial n_{20}}=-\mu_2+n_{20}g_2+\frac{\partial e_{LHY}}{\partial n_{20}}.
        \end{eqnarray}
   \begin{figure}[htbp]
   	\centering  
   	\subfigure[]{
   		\label{Fig.sub.1}
   		\includegraphics[width=0.23\textwidth]{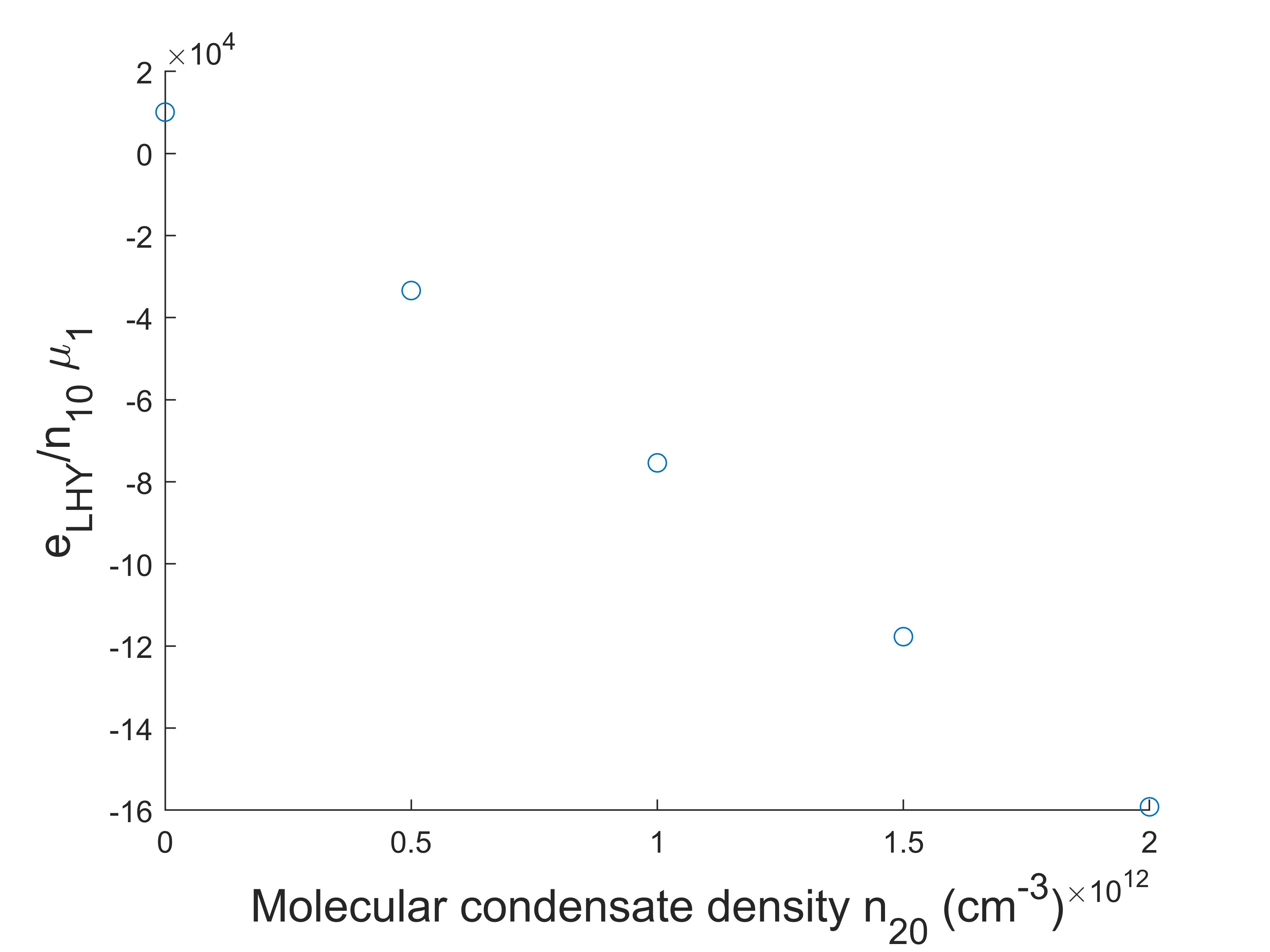}}
   	\subfigure[]{
   		\label{Fig.sub.2}
   		\includegraphics[width=0.23\textwidth]{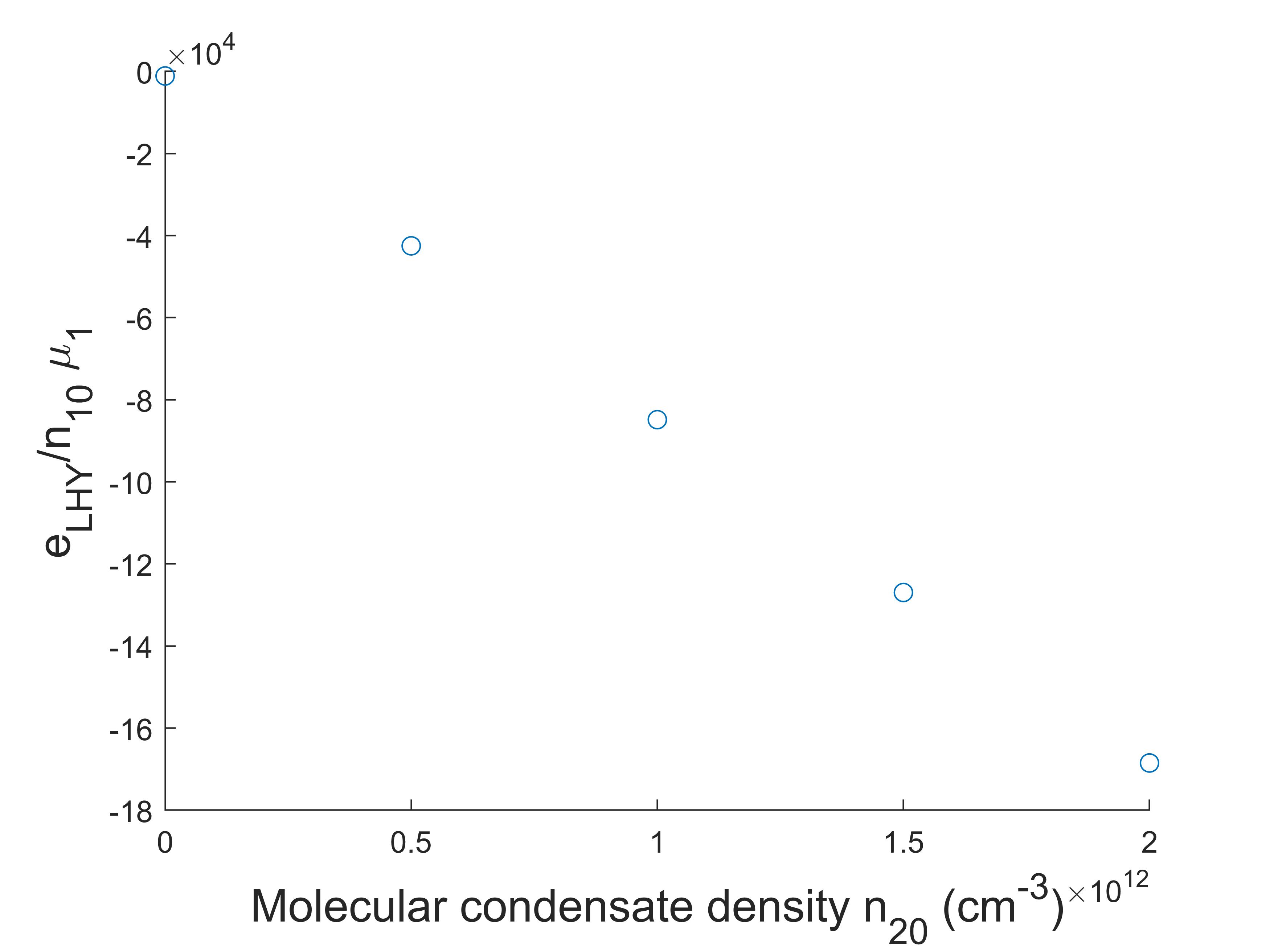}}
   		\subfigure[]{
   		\label{Fig.sub.3}
   		\includegraphics[width=0.23\textwidth]{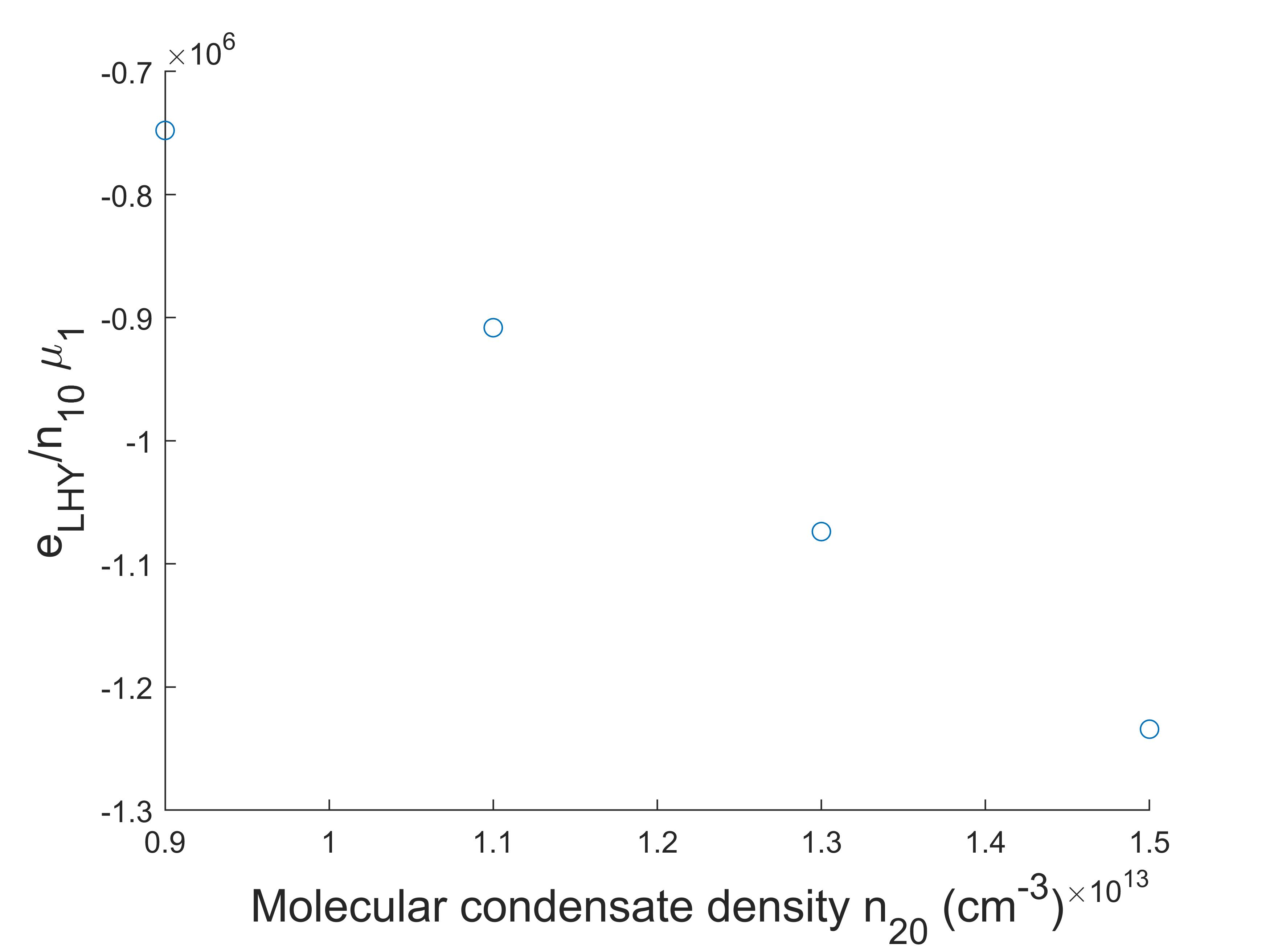}}
   	\subfigure[]{
   		\label{Fig.sub.4}
   		\includegraphics[width=0.23\textwidth]{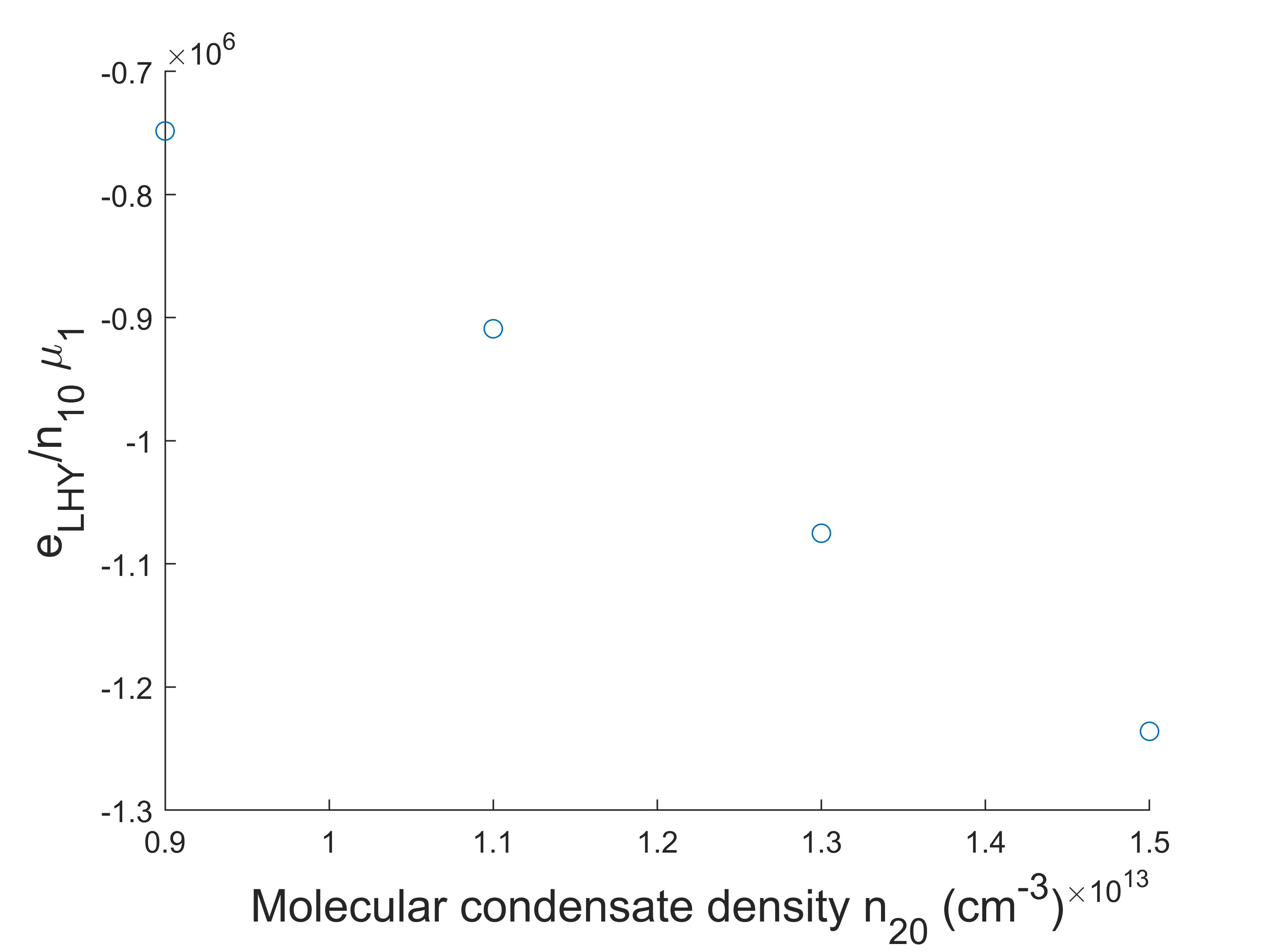}}
   	\caption{The LHY energy density $e_{LHY}$ as a function of molecular condensate density $n_{20}$ under different conditions. $\hbar=1,\,m=1,$ atomic scattering length $a_1=127a_0 (1-\frac{\Delta B}{B-B_0}),$ molecular scattering length $a_2=220a_0$, $a_0$ is the Bohr radius, $\Delta B$ is the resonance width,$\,  g=8.5\, \mu\mathrm{m}^{-\frac{1}{2}}.$ (a): close to ASF phase, $n_{10}=1.5\times 10^{13}\, \mathrm{cm}^{-3}\gg n_{20}$, $B-B_0=3\Delta B$. (b): close to ASF phase, $n_{10}=1.5\times 10^{13}\, \mathrm{cm}^{-3}\gg n_{20}$, $B-B_0=\frac{4}{3}\Delta B$. (c): close to MSF phase, $n_{10}=1.5\times 10^{11}\, \mathrm{cm}^{-3}\ll n_{20}$, $B-B_0=3\Delta B$. (d): close to MSF phase, $n_{10}=1.5\times 10^{11}\, \mathrm{cm}^{-3}\ll n_{20}$, $B-B_0=\frac{4}{3}\Delta B$.}
   	\label{alpha}
   \end{figure}
        $\frac{\partial e_{LHY}}{\partial n_{20}}$ is approximated as a negative constant $\alpha$ (see Fig. \ref{alpha}), which is independent of $n_{\sigma0}$ and $\nu$ . $\frac{\partial e_{LHY}}{\partial n_{10}}$ is much smaller than $n_{10}g_1$ so we negelect it. Using $n_0\equiv n_{10}+2n_{20}$ and assuming that $n_0$ is fixed, we obtain the molecular condensate density
        \begin{eqnarray}
        	n_{20}=\frac{-\alpha-\nu+2n_{0}g_1}{4g_1+g_2}\approx \frac{\abs{\alpha}-\abs{\mu_2}}{4g_1+g_2}.
        \end{eqnarray}
        As the magnetic field $B$ approaches the resonance position $B_0$ from the larger side, $\nu$ and $\abs{\mu_2}$ are decreasing (at the resonance position $\nu = 0$), and as $\abs{\mu_2}$ decreases to $\abs{\alpha}$, molecular condensate occurs and the critical detuning $\nu_c=\abs{\alpha}+2n_{0}g_1$, at which point the ASF phase undergoes a transition to the atomic-molecular superfluid (AMSF) and $\mu_2$ is still negative. As the detuning $\nu$ continues to decrease to $\nu_{c1}=\abs{\alpha}-\frac{1}{2}n_{0}g_2$ the molecular condenste density approaches total condenste density, at which point the AMSF phase undergoes a transition to the molecular superfluid (MSF). Within this beyond mean-field analysis, all transitions above are second order. We find $\abs{\alpha}\approx \Delta \mu\Delta B=k_{\mathrm{B}}\times410\mathrm{nK}$ when the coupling constant of the Feshbash resonance term $g=8.5\, \mu\mathrm{m}^{-\frac{1}{2}}$ and $\Delta\mu = h \times 770 \mathrm{kHz/G}$ is the relative magnetic moment\cite{zhang2020atomic}. $\abs{\alpha}$ is positively correlated with $g$ but when $g$ becomes too large, the excitation spectrum becomes unstable. In the AMSF phase, the ratio of atomic to molecular condensation $\frac{n_{10}}{n_{20}}=\frac{2\nu-2\abs{\alpha}+n_0g_2}{\abs{\alpha}-\nu+2n_0g_1}$, where experimentally $\abs{\alpha}$ can be solved from the precise ASF-AMSF transtion point $\nu_c=\abs{\alpha}+2n_{0}g_1$. These phase transtions are illustrated in Fig. \ref{tran}
    
        Symmetry-breaking in the phase transition also differs from the \textit{s}-wave case, in which the \textit{s}-wave Feshbach resonance interaction locks the phase of atomic condensate $\Psi_{1 0}=\abs{\Psi_{1 0}}e^{i\theta_1}$ and the phase of molecular condensate $\Psi_{2 0}=\abs{\Psi_{2 0}}e^{i\theta_2}$ together, with $\theta_2=2\theta_1(mod\, 2\pi)$\cite{radzihovsky2008superfluidity} and the symmetry of these two phases is broken at the same time in the Normal-AMSF phase transition. The \textit{g}-wave Feshbash resonance does not have this phase-locking relationship and thus the symmetry of the two phases is broken successively in different phase transitions: in the Normal-ASF transiton, $\theta_1$ is fixed and in the ASF-AMSF transtion, $\theta_2$ is fixed.
        
        \begin{acknowledgments}
        	FZ acknowledges support from the Innovation Program for Quantum Science and Technology under Grant No.~2021ZD0301904.
        \end{acknowledgments}
        \bibliographystyle{unsrt}
        \bibliography{molecularbec}
\begin{widetext}
\section*{Supplementary Materials}
\subsection*{I}
The Feshbach resonance term in (\ref{ham}) is obtained from a more general Hamiltonian:
\begin{equation}
	H_{FR}=\int d\vb r d\vb r' [g(\vb r-\vb r')\Phi(\vb r, \vb r')\Psi^{\dag}(\vb r)\Psi^{\dag}(\vb r')+h.c.]. \label{fr}
\end{equation}
Here $\Psi(\vb r)\equiv \sum_{\vb p}e^{i\vb p\cdot \vb r}c_{\vb p}$ is a boson field operator, and
\begin{equation}
	\Phi(\vb r, \vb r')\equiv\sum\limits_{\vb q}e^{i\vb q\cdot \vb R}\sum\limits_{n,l,m}u_{nl}(\tilde r)Y_{l,m}(\vb{\tilde r})b_{\vb q,n,l,m} \label{phi}
\end{equation}
describes molecules with center of mass $\vb R\equiv(\vb r + \vb r')/2$, and relative coordinate $\vb{\tilde r}\equiv\vb r - \vb r'$. $b_{\vb q,n,l,m}$ is an annihilation operator of a bound molecular state, described by the eigenfunction $u_{nl}(\tilde r)Y_{l,m}(\vb{\tilde r})$. Substituting (\ref{phi}) into (\ref{fr}), one obtains
\begin{eqnarray}
	&H_{FR}=\int d\vb r d\vb r' [g(\vb r-\vb r')\sum\limits_{\vb q}e^{i\vb q\cdot \vb R}\sum\limits_{n,m}u_{nl}(\tilde r)Y_{l,m}(\vb{\tilde r})b_{\vb q,n,l,m}\sum\limits_{\vb p}e^{-i\vb p\cdot \vb r}c^{\dag}_{\vb p}\sum\limits_{\vb p'}e^{-i\vb p'\cdot \vb r'}c^{\dag}_{\vb p'}+h.c.]\\ \notag 
	&=\int d\vb{\tilde r} d\vb R [g(\tilde r)\sum\limits_{\vb q}e^{i\vb q\cdot \vb R}\sum\limits_{n,m}u_{nl}(\tilde r)Y_{l,m}(\vb{\tilde r})b_{\vb q,n,l,m}\sum\limits_{\vb p}e^{-i\vb p\cdot (\vb R+\frac{\vb{\tilde r}}{2})}c^{\dag}_{\vb p}\sum\limits_{\vb p'}e^{-i\vb p'\cdot (\vb R-\frac{\vb{\tilde r}}{2})}c^{\dag}_{\vb p'}+h.c.]\\ \notag
	&=\int d\vb{\tilde r}[g(\tilde r)\sum\limits_{\vb{q,p,p'}}\delta_{\vb{q,p+p'}}\sum\limits_{n,m}u_{nl}(\tilde r)Y_{l,m}(\vb{\tilde r})b_{\vb q,n,l,m}e^{-i\frac{\vb{\tilde r}}{2}(\vb{p-p'})}c^{\dag}_{\vb p}c^{\dag}_{\vb p'}+h.c.]\\ \notag
	&=\int d\vb{\tilde r}[g(\tilde r)\sum\limits_{\vb{q,p}}\sum\limits_{n,m}u_{nl}(\tilde r)Y_{l,m}(\vb{\tilde r})b_{\vb q,n,l,m}e^{-i\vb{{\tilde r}\cdot p}}c^{\dag}_{\vb{p+\frac{q}{2}}}c^{\dag}_{\vb{-p+\frac{q}{2}}}+h.c.] , \label{frt}
\end{eqnarray}
where
\begin{eqnarray}
	&e^{-i\vb{r\cdot p}}=e^{-i\abs{\vb r}\abs{\vb p} cos\theta}=\sum\limits_{l=0}^{\infty}(-i)^l(2l+1)j_l(\abs{\vb r}\abs{\vb p})P_l(cos\theta)\\
	&=\sum\limits_{l=0}^{\infty}\sum\limits_{m=-l}^{l}(-i)^l\frac{4\pi}{2l+1}(2l+1)j_l(\abs{\vb r}\abs{\vb p})Y_{lm}(\hat{\vb r})Y^*_{lm}(\hat{\vb p}) \label{e}
\end{eqnarray}
and	$P_l(\vb{\hat r\cdot \hat p})=\frac{4\pi}{2l+1}\sum\limits_{m=-l}^{l}Y_{lm}(\hat{\vb r})Y^*_{lm}(\hat{\vb p})$ is the Legendre polynomial.
For specific $l$ and $m$, substituting (\ref{e}) into (\ref{frt}), one obtains
\begin{eqnarray}
	&H_{FR}=\int d\vb{\tilde r}[g(\tilde r)\sum\limits_{\vb{q,p,n}}u_{nl}(\tilde r)Y_{l,m}(\vb{\tilde r})b_{\vb q,n,l,m}e^{-i\vb{{\tilde r}\cdot p}}c^{\dag}_{\vb{p+\frac{q}{2}}}c^{\dag}_{\vb{-p+\frac{q}{2}}}+h.c.]\\ \notag
	&=\int d\vb{\tilde r}g(\tilde r)\sum\limits_{\vb{q,p,n}}u_{nl}(\tilde r)Y_{l,m}(\vb{\tilde r})b_{\vb q,n,l,m}
	\\ \notag
	&\times \sum\limits_{l'=0}^{\infty}\sum\limits_{m'=-l'}^{l'}[(-i)^{l'}\frac{4\pi}{2l'+1}(2l'+1)j_{l'}(\abs{\vb {\tilde r}} \abs{\vb p})Y_{l'm'}(\hat{{\vb {\tilde r}}})Y^*_{l'm'}(\hat{\vb p})]c^{\dag}_{\vb{p+\frac{q}{2}}}c^{\dag}_{\vb{-p+\frac{q}{2}}}+h.c.\\ \notag
	&=	\sum\limits_{\vb{q,p}}g_pj_l(\abs{\vb {\tilde r_0}}\abs{\vb p})Y^*_{l,m}(\hat{\vb p})b_{\vb q}c^{\dag}_{\vb{p+\frac{q}{2}}}c^{\dag}_{\vb{-p+\frac{q}{2}}}+h.c..
\end{eqnarray}
$j_l(\abs{\vb {\tilde r_0}}\abs{\vb p})$ is the spherical Bessel function, which is approximately $p^l$ in small $p$ ($p<\frac{1}{\tilde r_0}$) and converges as $p$ becomes larger. The order of magnitude of $\abs{\vb {\tilde r_0}}$ is the size of molecule.
For our g wave case, $l=4, m=2$,

\begin{eqnarray}
	H_{FR}=\sum\limits_{\vb{q,p}}g_pj_4(\abs{\vb {\tilde r_0}}\abs{\vb p})Y^*_{42}(\hat{\vb p})b_{\vb q}c^{\dag}_{\vb{p+\frac{q}{2}}}c^{\dag}_{\vb{-p+\frac{q}{2}}}+h.c..
\end{eqnarray}
\subsection*{II} 
The LHY energy density is
\begin{eqnarray}
	e_{LHY}=\frac{1}{2V}\sum\limits_{\vb k\sigma}(E_{\vb{k}\sigma}-\tilde{\varepsilon}_{\vb k \sigma}+\frac{(n_{\sigma 0}g_{\sigma})^2}{2{\varepsilon}_{\vb k \sigma}}) ,
\end{eqnarray}
and in the quasi-2D case $k_z = n\pi/L (n = 0, \pm1, ...)$. Replace summation with integration, one obtains
\begin{eqnarray}
	&e_{LHY}=\frac{1}{2(2\pi)^2 2L}\int d^2 k \sum\limits_{n \sigma}(E_{n\vb{k}\sigma}-\tilde{\varepsilon}_{n\vb{k}})+\frac{1}{2(2\pi)^3}\int d^3 k \sum\limits_{\sigma}\frac{(n_{\sigma 0}g_{\sigma})^2}{2{\varepsilon}_{\vb k \sigma}}\\  
	&\equiv\frac{1}{2(2\pi)^2 l}\int d^2 k'\sum\limits_{n \sigma}(E'_{n\vb{k}\sigma}-\tilde{\varepsilon}'_{n\vb{k}})+\frac{1}{2(2\pi)^3}\int d^3 k' \sum\limits_{\sigma}\frac{(g'_{\sigma})^2}{2{\varepsilon}'_{\vb k \sigma}}, \label{elhy}
\end{eqnarray}
where we define
\begin{equation}
	l=2L/l_0, k'=k/(1/l_0),{\varepsilon}'_{\vb k \sigma}=k'^2/2m_{\sigma},m_1=1,m_2=2,E'=El_0^2,g'_{\sigma}=n_{\sigma 0}g_{\sigma}l_0^2,x=k'^2+(\frac{2\pi}{l})^2n^2.
\end{equation}
The first integration in (\ref{elhy}) is
\begin{eqnarray}
	&\frac{1}{2(2\pi)^2 l}\int d^2 k'\sum\limits_{n \sigma}(E'_{n\vb{k}\sigma}-\tilde{\varepsilon}'_{n\vb{k}})=\frac{1}{2(2\pi)^2 l}\int  k' d k'd\phi\sum\limits_{n \sigma}(E'_{n\vb{k}\sigma}-\tilde{\varepsilon}'_{n\vb{k}})\\
	&=	\frac{1}{2(2\pi)^2 l}\int_{0}^{\Lambda} k' d k'd\phi\sum\limits_{n \sigma}(E'_{n\vb{k}\sigma}-\tilde{\varepsilon}'_{n\vb{k}})+
	\frac{1}{2(2\pi)^2 l}\int_{\Lambda}^{\Lambda_1} k' d k'd\phi\sum\limits_{n \sigma}(E'_{n\vb{k}\sigma}-\tilde{\varepsilon}'_{n\vb{k}}),\,	\Lambda_1\to \infty,\,\Lambda\gg l_0/\abs{\vb {\tilde r_0}} .
\label{i1}
\end{eqnarray}
The last integration in (\ref{i1}) is
\begin{eqnarray}
	&\frac{1}{2(2\pi)^2 l}\int_{\Lambda}^{\Lambda_1} k' d k'd\phi\sum\limits_{n \sigma}(E'_{n\vb{k}\sigma}-\tilde{\varepsilon}'_{n\vb{k}})
	=	\frac{1}{2(2\pi) l}\int_{\Lambda}^{\Lambda_1} k' d k'\sum\limits_{n \sigma}(\sqrt{\tilde{\varepsilon}'^2_{n\vb{k}\sigma}-g'^2_{\sigma}}-\tilde{\varepsilon}'_{n\vb{k}\sigma})\\ \notag
	&\approx	\frac{1}{2(2\pi) l}\int_{\Lambda}^{\Lambda_1} k' d k'\sum\limits_{n \sigma}(\frac{-g'^2_{\sigma}}{x/\sigma}), 
\end{eqnarray}
when $n=0$
\begin{equation}
	\frac{1}{2(2\pi) l}\int_{\Lambda}^{\Lambda_1} k' d k'\sum\limits_{\sigma}(\frac{-g'^2_{\sigma}}{k'^2/\sigma})=\sum\limits_{\sigma}\frac{-\sigma g'^2_{\sigma}}{2(2\pi) l}\int_{\Lambda}^{\Lambda_1} d k'\frac{1}{k'}=\sum\limits_{\sigma}\frac{-\sigma g'^2_{\sigma}}{(4\pi) l}\ln(\frac{\Lambda_1}{\Lambda}),
\end{equation}
and when $n\neq0$
\begin{eqnarray}
	&\frac{1}{2(2\pi) l}\int_{\Lambda}^{\Lambda_1} k' d k'\sum'\limits_{n \sigma}(\frac{-g'^2_{\sigma}}{x/\sigma})=\frac{1}{(4\pi) l}\sum'\limits_{n \sigma}-\sigma g'^2_{\sigma}\int_{\Lambda}^{\Lambda_1}  d k'\frac{k'}{k'^2+(\frac{2\pi}{l})^2n^2}=\frac{1}{4\pi l}\sum\limits_{ \sigma}-\sigma g'^2_{\sigma}\int_{\Lambda}^{\Lambda_1}  d k'(\frac{l}{2}\coth{\frac{k'l}{2}}-\frac{1}{k'}) \\ \notag
	&\approx \frac{1}{4\pi l}\sum\limits_{ \sigma}-\sigma g'^2_{\sigma}\int_{\Lambda}^{\Lambda_1}  d k'(\frac{l}{2}-\frac{1}{k'})=	\frac{1}{4\pi l}\sum\limits_{ \sigma}-\sigma g'^2_{\sigma}[\frac{l}{2}(\Lambda_1-\Lambda)-\ln (\frac{\Lambda_1}{\Lambda})] \\ \notag
	&=\sum\limits_{ \sigma}	\frac{-\sigma g'^2_{\sigma}}{8\pi }\Lambda_1+\sum\limits_{ \sigma}	\frac{\sigma g'^2_{\sigma}}{8\pi }\Lambda+\sum\limits_{ \sigma}	\frac{\sigma g'^2_{\sigma}}{4\pi l}\ln (\frac{\Lambda_1}{\Lambda}).
\end{eqnarray}
The last integration in (\ref{elhy}) is
\begin{equation}
	\frac{1}{2(2\pi)^3}\int d^3 k' \sum\limits_{\sigma}\frac{(g'_{\sigma})^2}{2{\varepsilon}'_{\vb k \sigma}}=	\frac{1}{2(2\pi)^3}\int d^3 k' \sum\limits_{\sigma}\frac{\sigma(g'_{\sigma})^2}{k'^2}=\sum\limits_{ \sigma}	\frac{\sigma g'^2_{\sigma}}{8\pi }\Lambda_1
\end{equation}
The LHY energy density is 
\begin{equation}
	e_{LHY}
	=\frac{1}{2(2\pi)^2 l}\int_{0}^{\Lambda} k' d k'\int_{0}^{2\pi}d\phi\sum\limits_{n \sigma}(E'_{n\vb{k'}\sigma}-\tilde{\varepsilon}'_{n\vb{k'}})+\sum\limits_{ \sigma}	\frac{\sigma g'^2_{\sigma}}{8\pi }\Lambda,
\end{equation}
which can be calculated numerically.
 \end{widetext}          
       \end{document}